\documentclass[twocolumn,tighten,trackchanges]{aastex61}
\usepackage{natbib}
\usepackage{pslatex}
\usepackage{graphicx}
\newcommand{\Htwoo}{\mbox{H$_{2}$O}}

\newcommand{\cotwo}{\mbox{CO$_{2}$}}
\newcommand{\otwo}{\mbox{O$_{2}$}}

\newcommand{\xsig}{$X\,^1\Sigma ^{+}$}
\newcommand{\api}{$A\,^1\Pi$}
\newcommand{\atpi}{$a\,^3\Pi$}

\newcommand{\csig}{$C\,^1\Sigma ^{+}$}

\newcommand{\kms}{km~s$^{-1}$}

\shorttitle{FUV Spectral Signatures}
\shortauthors{Feldman, et al.}

\begin{document}

\title{FUV Spectral Signatures of Molecules and the Evolution of the Gaseous Coma of Comet 67P/Churyumov-Gerasimenko} 

\correspondingauthor{Paul D. Feldman}
\email{pfeldman@jhu.edu}

\author[0000-0002-9318-259X]{Paul D. Feldman}
\affil{Department of Physics and Astronomy, The Johns Hopkins University, 3400 N. Charles Street, Baltimore, Maryland 21218}

\author{Michael F. A'Hearn}
\altaffiliation{deceased}
\affil{Astronomy Department, University of Maryland, College Park, MD 20742}

\author{Jean-Loup Bertaux}
\affil{LATMOS, CNRS/UVSQ/IPSL, 11 Boulevard d'Alembert, 78280 Guyancourt, France}

\author{Lori M. Feaga}
\affil{Astronomy Department, University of Maryland, College Park, MD 20742}

\author{Brian A. Keeney}
\affil{Southwest Research Institute, Department of Space Studies, Suite 300, 1050 Walnut Street, Boulder, CO 80302}

\author{Matthew M. Knight}
\affil{Astronomy Department, University of Maryland, College Park, MD 20742}

\author{John Noonan}
\affil{Southwest Research Institute, Department of Space Studies, Suite 300, 1050 Walnut Street, Boulder, CO 80302}

\author{Joel Wm. Parker}
\affil{Southwest Research Institute, Department of Space Studies, Suite 300, 1050 Walnut Street, Boulder, CO 80302}

\author{Eric Schindhelm}
\affil{Ball Aerospace and Technology Corp, 1600 Commerce Street, Boulder, Colorado 80301}

\author{Andrew J. Steffl}
\affil{Southwest Research Institute, Department of Space Studies, Suite 300, 1050 Walnut Street, Boulder, CO 80302}

\author{S. Alan Stern}
\affil{Southwest Research Institute, Department of Space Studies, Suite 300, 1050 Walnut Street, Boulder, CO 80302}

\author{Ronald J. Vervack}
\affil{Space Exploration Sector, Johns Hopkins University Applied Physics Laboratory, 11100 Johns Hopkins Road, Laurel, MD 20723-6099}

\author{Harold A. Weaver}
\affil{Space Exploration Sector, Johns Hopkins University Applied Physics Laboratory, 11100 Johns Hopkins Road, Laurel, MD 20723-6099}

\pagestyle{myheadings}
\markright{\today}

\begin{abstract}

The Alice far-ultraviolet imaging spectrograph onboard {\it Rosetta} observed emissions from atomic and molecular species from within the coma of comet 67P/Churyumov-Gerasimenko during the entire escort phase of the mission from 2014 August to 2016 September.  The initial observations showed that emissions of atomic hydrogen and oxygen close to the surface were produced by energetic electron impact dissociation of \Htwoo.  Following delivery of the lander, {\it Philae}, on 2014 November 12, the trajectory of {\it Rosetta} shifted to near-terminator orbits that allowed for these emissions to be observed against the shadowed nucleus that, together with the compositional heterogeneity, enabled us to identify unique spectral signatures of dissociative electron impact excitation of \Htwoo, \cotwo, and \otwo.  CO emissions were found to be due to both electron and photoexcitation processes.  Thus we are able, from far-ultraviolet spectroscopy, to qualitatively study the evolution of the primary molecular constituents of the gaseous coma from start to finish of the escort phase.  Our results show asymmetric outgassing of \Htwoo\ and \cotwo\ about perihelion, \Htwoo\ dominant before and \cotwo\ dominant after, consistent with the results from both the {\it in situ} and other remote sensing instruments on {\it Rosetta}.

\end{abstract}


\section{INTRODUCTION}

We have previously \citep{Feldman:2015} described the initial observations of the near-nucleus coma of comet 67P/Churyumov-Gerasimenko made by the Alice far-ultraviolet imaging spectrograph onboard {\it Rosetta} in the first few months following orbit insertion in August 2014. These observations of the sunward limb, made from distances between 10 and 30 km from the comet's nucleus, showed emissions of atomic hydrogen, oxygen, and carbon, spatially localized within a few~km of the nucleus and attributed to electron impact dissociation of \Htwoo\ and \cotwo\ vapor.  This interpretation was supported by measurements of suprathermal electrons by the Rosetta Plasma Consortium (RPC) instruments on {\it Rosetta} \citep{Clark:2015,Galand:2016}.

In our initial report we noted that following lander delivery {\it Rosetta} shifted to a $\sim$30~km near-terminator orbit that enabled observations of the illuminated coma along a short line-of-sight to shadowed regions of the nucleus.  In addition to the high solar phase angle associated with this orbit that reduced the light from illuminated regions of the nucleus, this geometry also removed contributions to the observed atomic emissions from the extended coma and, in the case of \ion{H}{1}, the interplanetary emissions.  Under these observing conditions, the observed relative intensities of \ion{H}{1} Lyman-$\beta$, Lyman-$\alpha$, \ion{O}{1} $\lambda$1304, and \ion{O}{1} $\lambda$1356, are found to be consistent with the laboratory spectra produced by electron impact dissociation of \Htwoo\ \citep{Makarov:2004}, confirming this mechanism as the source of the cometary emissions.  Based on coma models \citep{Combi:2004}, we calculate that solar resonance scattering by H and O atoms along the line-of-sight contribute negligibly to the observed emissions.
 
In this paper we examine several other observations made against the shadowed nucleus at relatively close range both before and after perihelion that exhibited unique spectral signatures of other coma molecules, \cotwo, CO, and \otwo.  These are then used to interpret limb observations to investigate the longer term evolution of the coma of 67P over the entire escort phase of the {\it Rosetta} mission, up to the penultimate mission day, 2016 September 29.  Our analysis complements and confirms the results of several other teams \citep{Bockelee:2016,Biver:2016,Hansen:2016,Gasc:2017,Hoang:2017}.

\section{OBSERVATIONS}

\subsection{Instrument Description \label{inst}}

The Alice instrument has been described in detail previously \citep{Stern:2007}.  It is a lightweight, low-power, imaging spectrograph designed for in situ far-ultraviolet imaging spectroscopy of comet 67P in the spectral range 700-2050~\AA.  The slit is in the shape of a dog bone, 5.5\degr\ long, with a width of 0.05\degr\ in the central 2.0\degr, while the ends are 0.10\degr\ wide, giving a spectral resolution between 8 and 12~\AA\ for extended sources that fill its field-of-view.  Each spatial pixel or row along the slit is 0.30\degr\ long.  All of the spectra presented in this paper are averages over portions of the narrow center region of the slit, providing the best spectral resolution possible with Alice.

To maintain the sensitivity of the Alice detector over the course of a long mission, the detector photocathode was configured to minimize the effect of prolonged exposure to \ion{H}{1} Lyman-$\alpha$ radiation.  This was done by leaving the detector uncoated in a strip the width of the image of the spectrograph slit at the position of Lyman-$\alpha$, 1216~\AA.  The detector was coated on either side of this strip, with CsI on the long wavelength side and KBr at short wavelengths \citep{Stern:2007}.  Despite this precaution, there was significant degradation of the performance at Lyman-$\alpha$ over the escort phase leading to uncertainties in the observed spectral shape and absolute flux of the observed Lyman-$\alpha$ emission.  Another effect is that long wavelength light from the nucleus, scattered internally in the instrument, gives rise to uniform background signals (``red leak'') that differ on either side of Lyman-$\alpha$.  As these vary with both time and viewing geometry, no attempt is made to remove this background in the pipeline processing.  The Alice spectrograph operated successfully until the very end of the mission on 2016 September 30.

\subsection{Observations Toward the Shadowed Nucleus}

\begin{figure*}[h]
\begin{center}
\includegraphics*[width=0.37\textwidth,angle=0.]{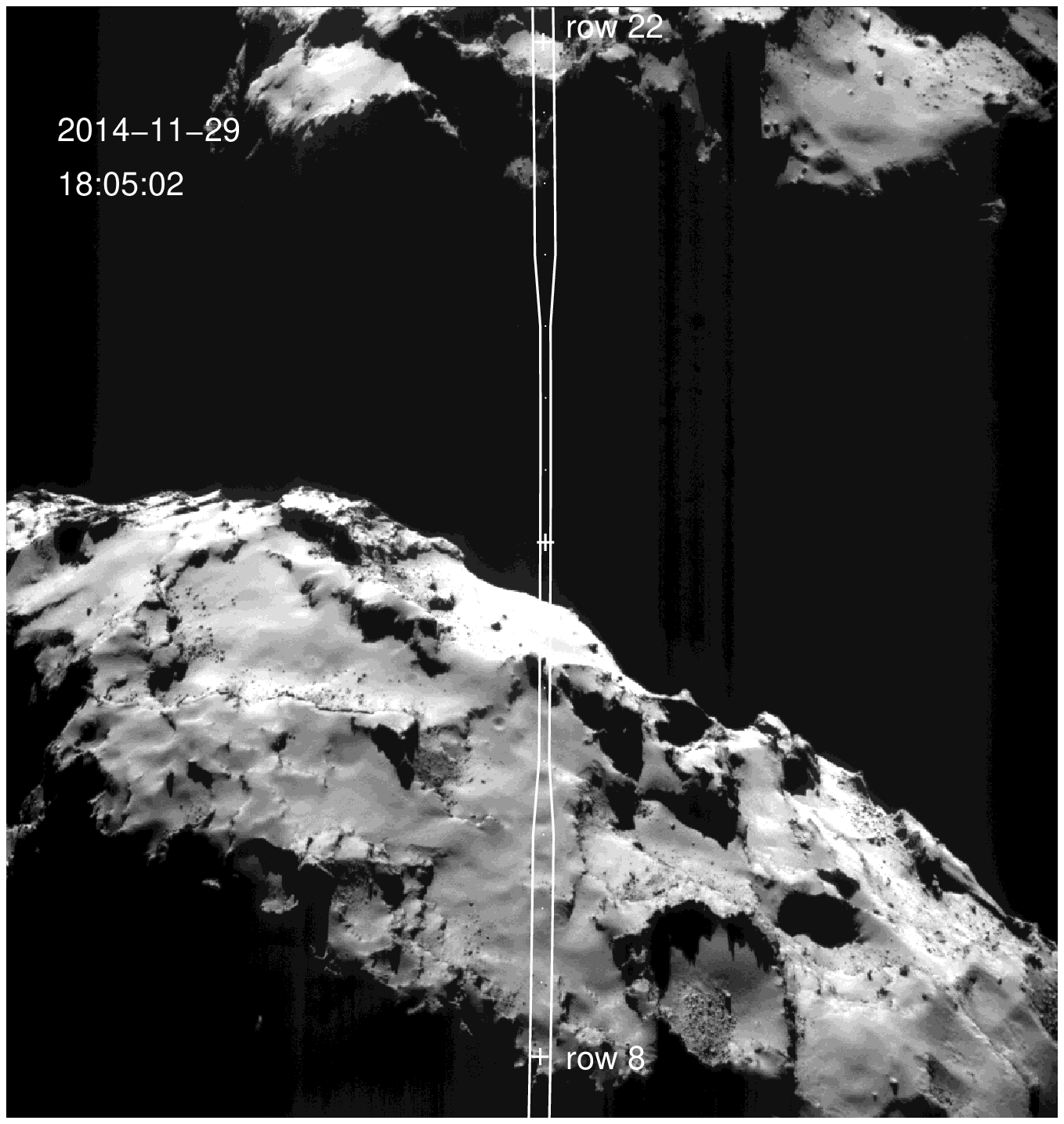}
\includegraphics*[width=0.62\textwidth,angle=0.]{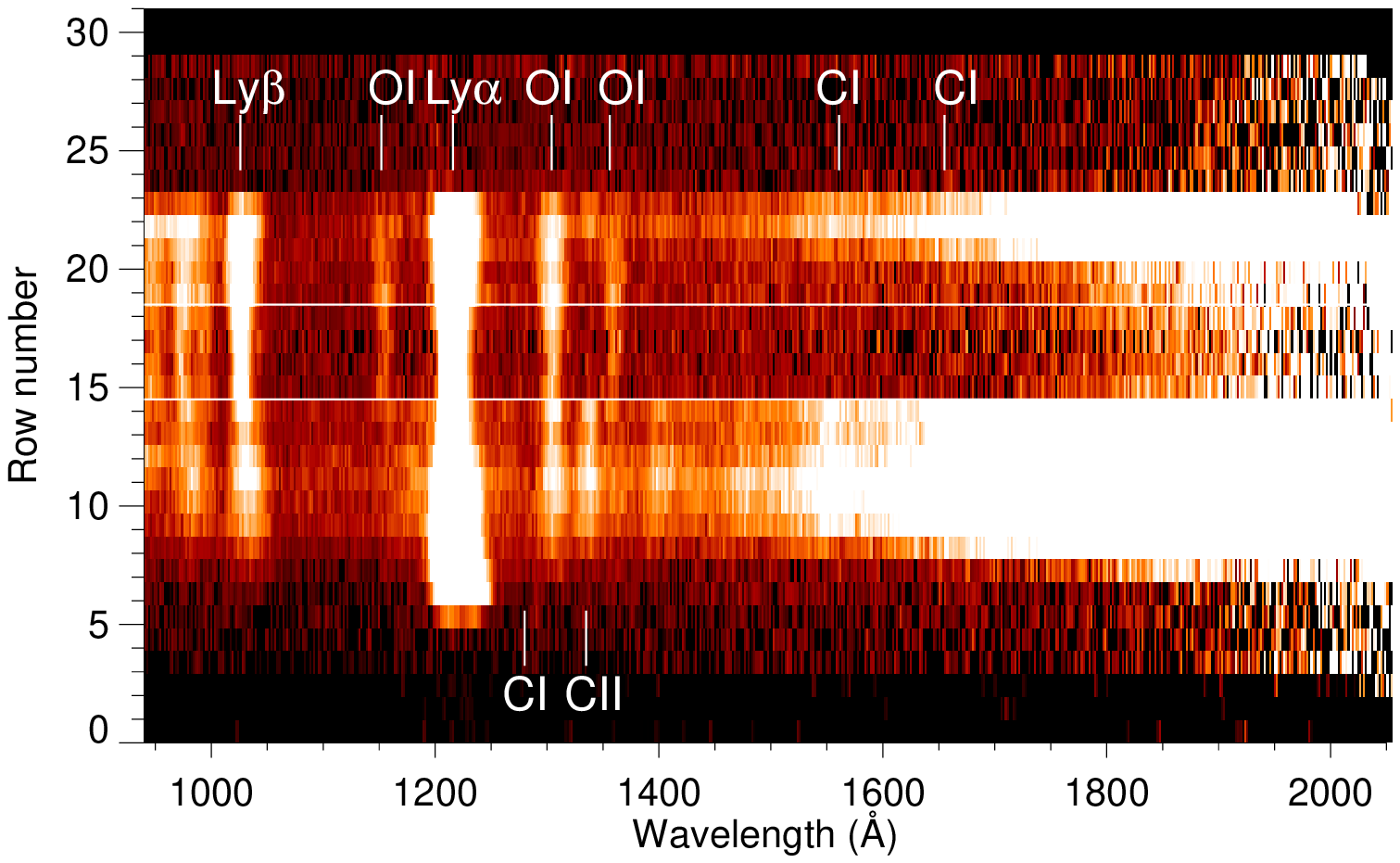}
\includegraphics*[width=0.37\textwidth,angle=0.]{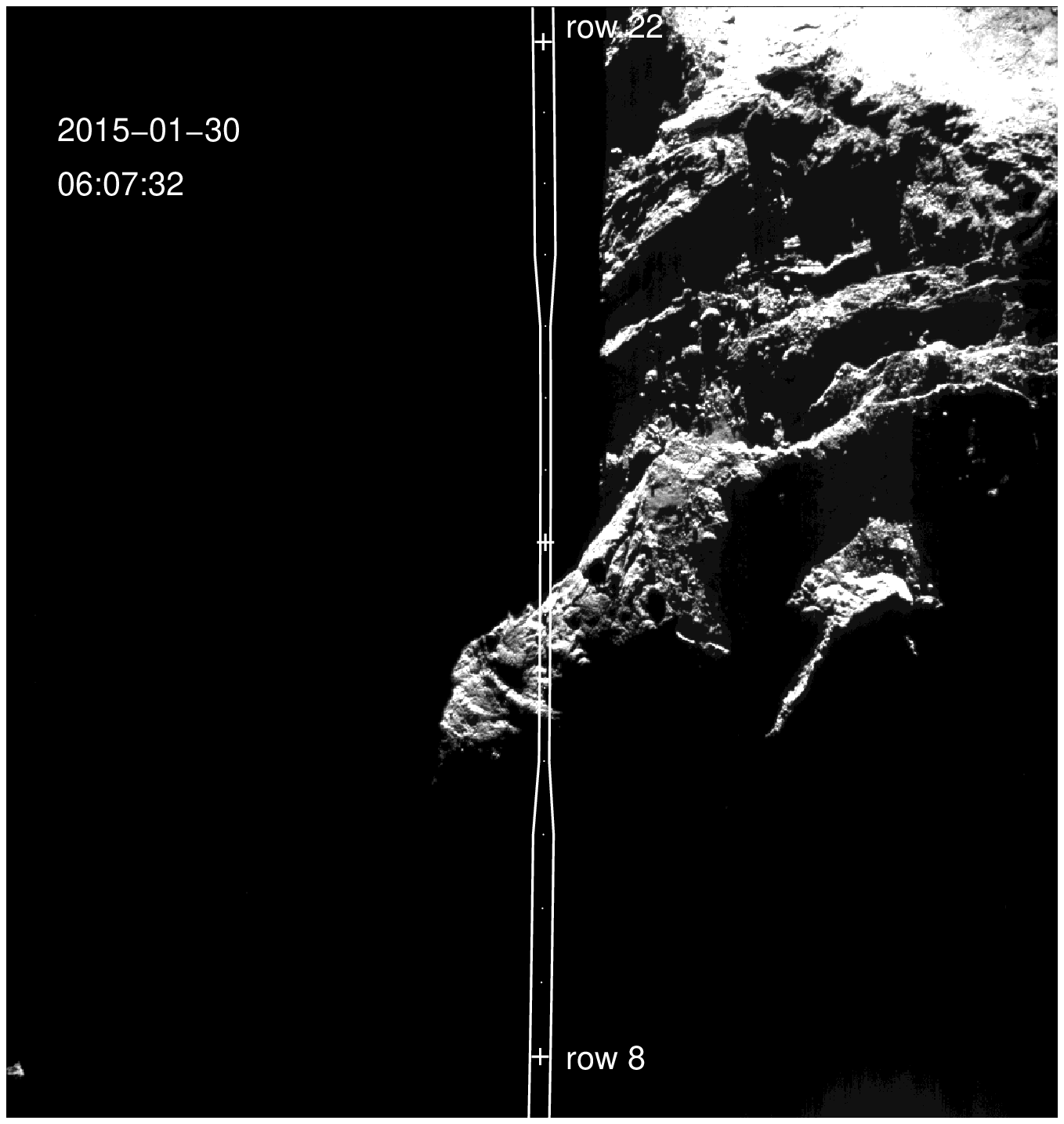}
\includegraphics*[width=0.62\textwidth,angle=0.]{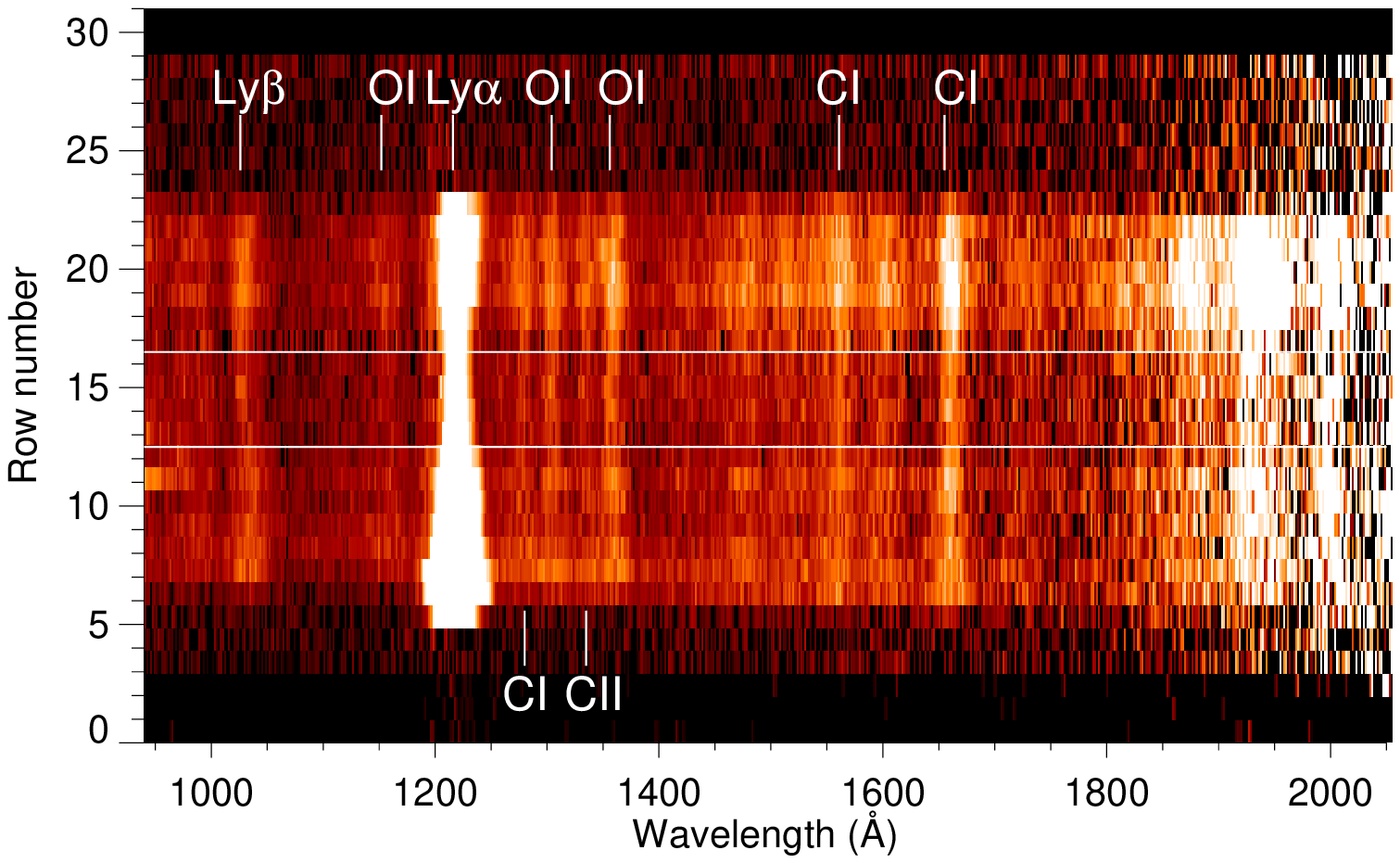}
\includegraphics*[width=0.37\textwidth,angle=0.]{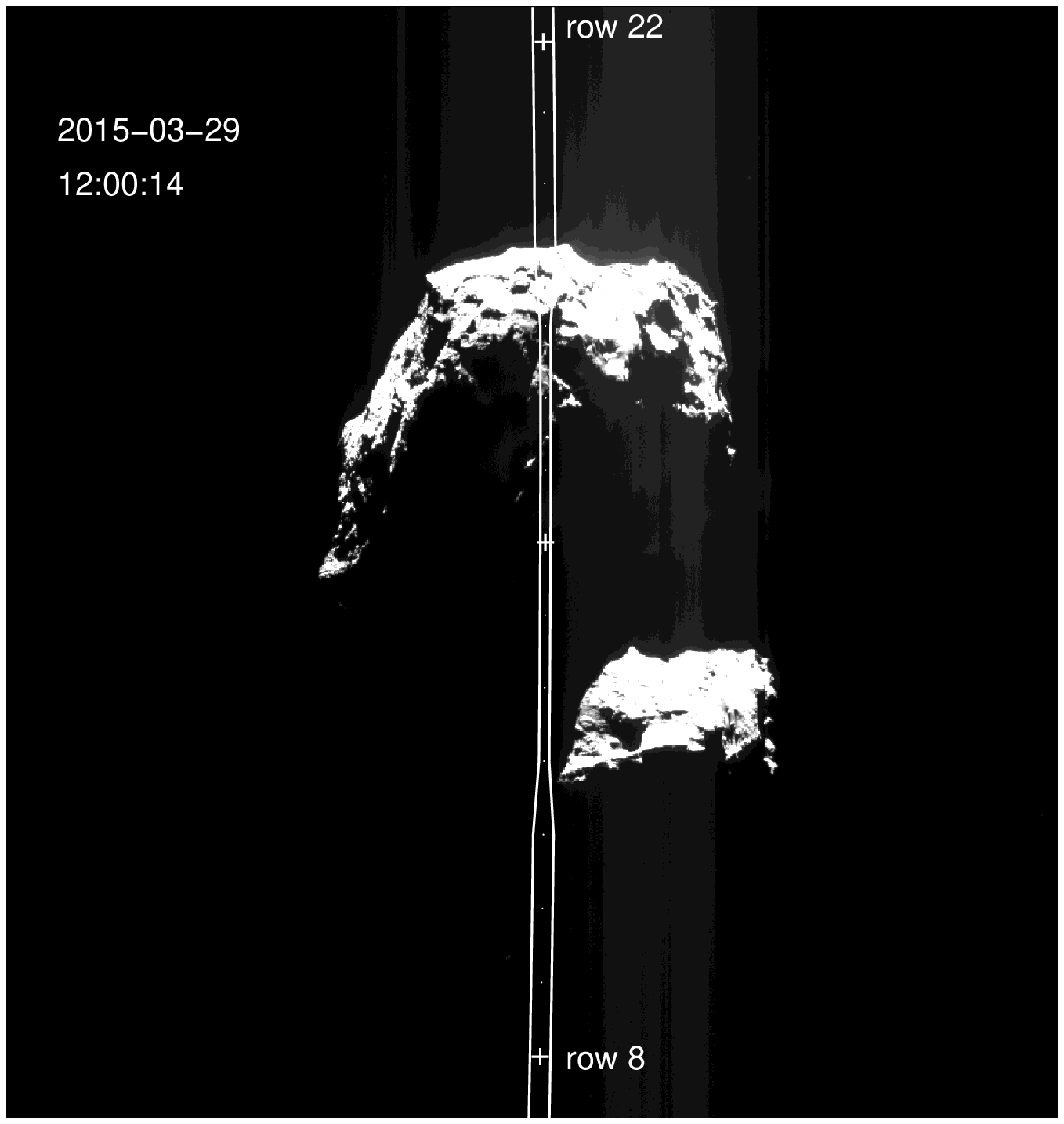}
\includegraphics*[width=0.62\textwidth,angle=0.]{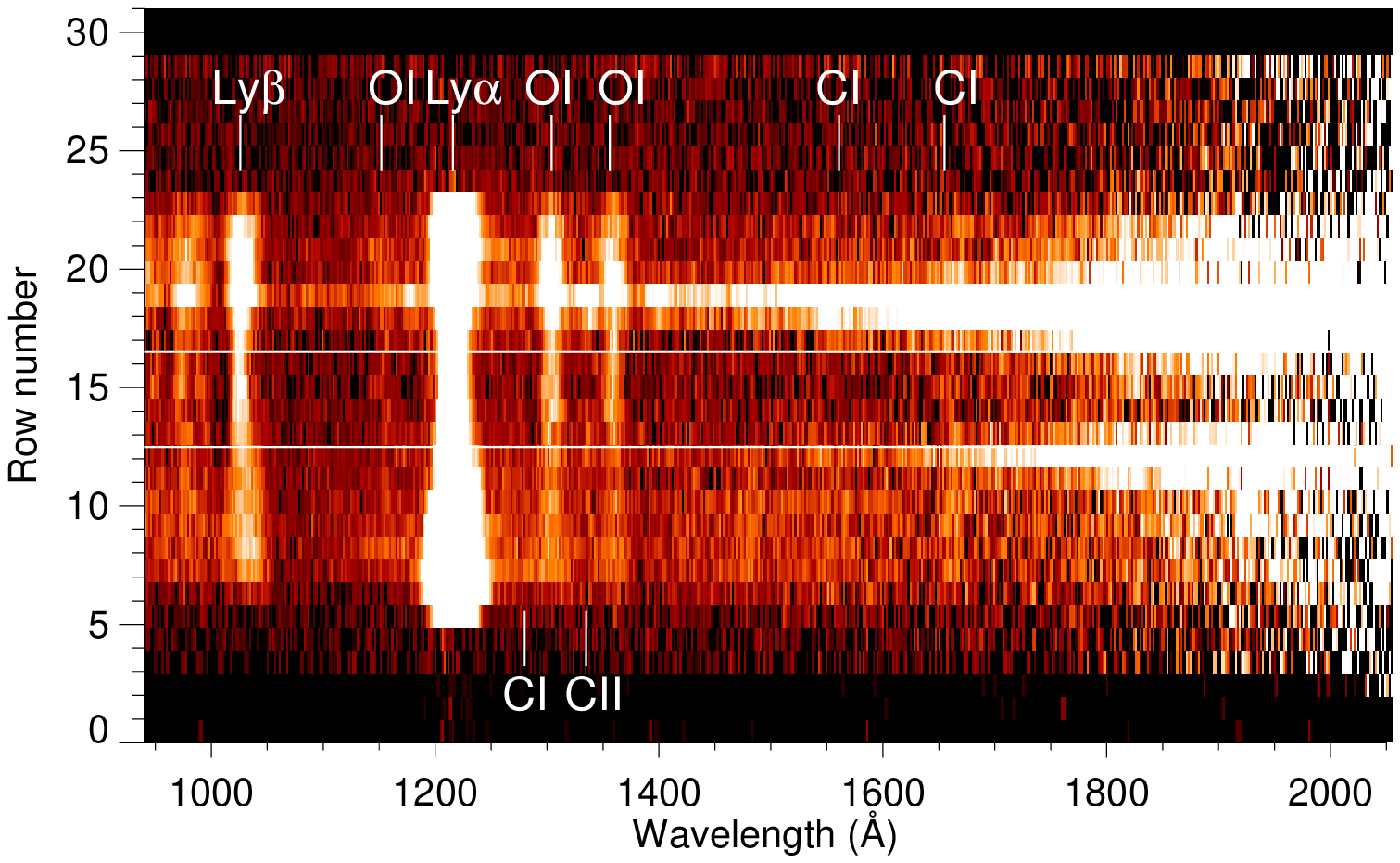}
\caption{NAVCAM context images (left) with the Alice slit superimposed and near simultaneous spectral images (right) for the three pre-perihelion dates given in Table~\ref{obs}.  Observation parameters are given in the table.  The white horizontal lines outline the 4 rows used in the spectral extraction.  In all of the images the Sun is towards the top.  \label{nav1} }
\end{center}
\end{figure*}

\begin{figure}[h]
\begin{center}
\includegraphics*[width=0.44\textwidth,angle=0.]{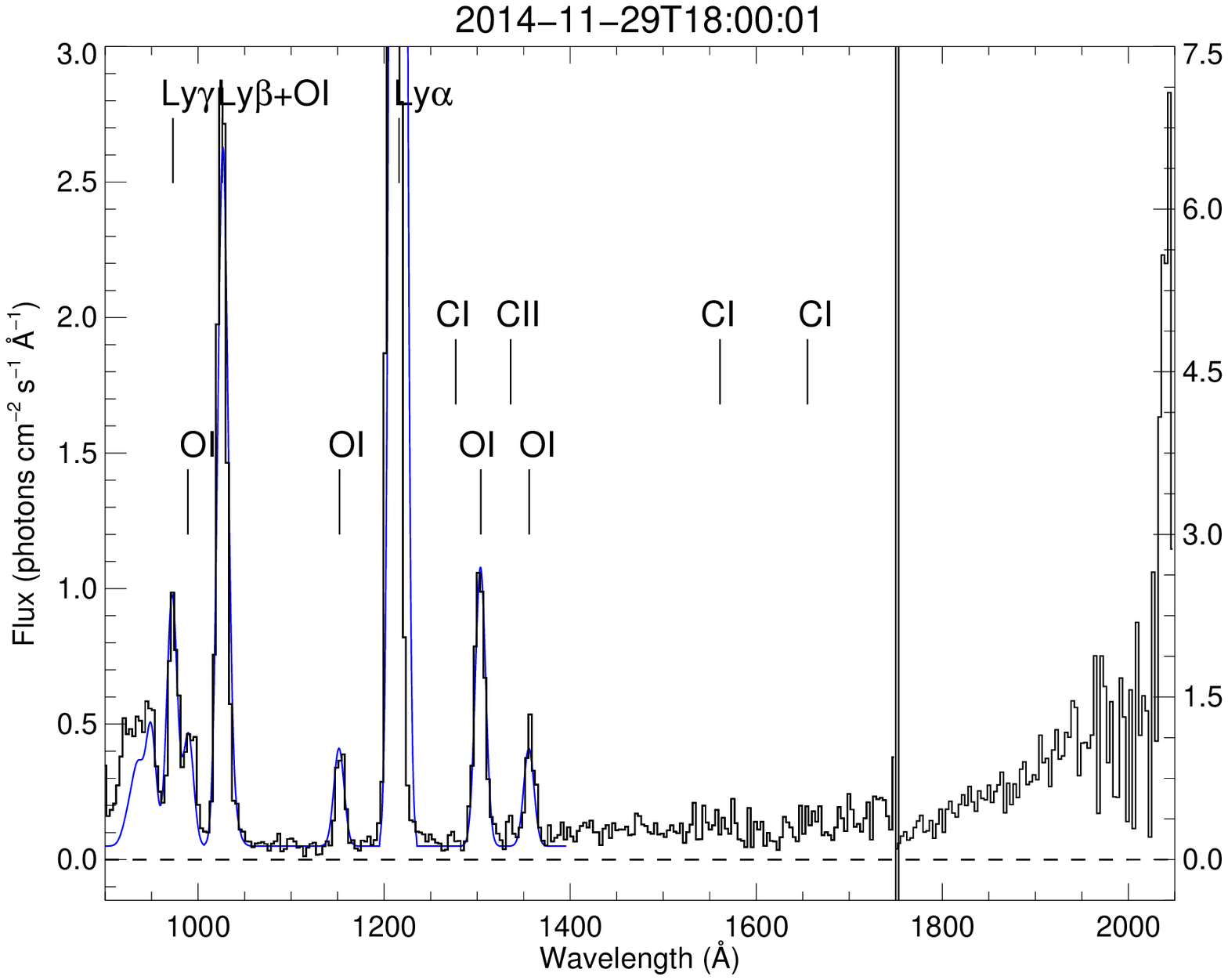}
\includegraphics*[width=0.44\textwidth,angle=0.]{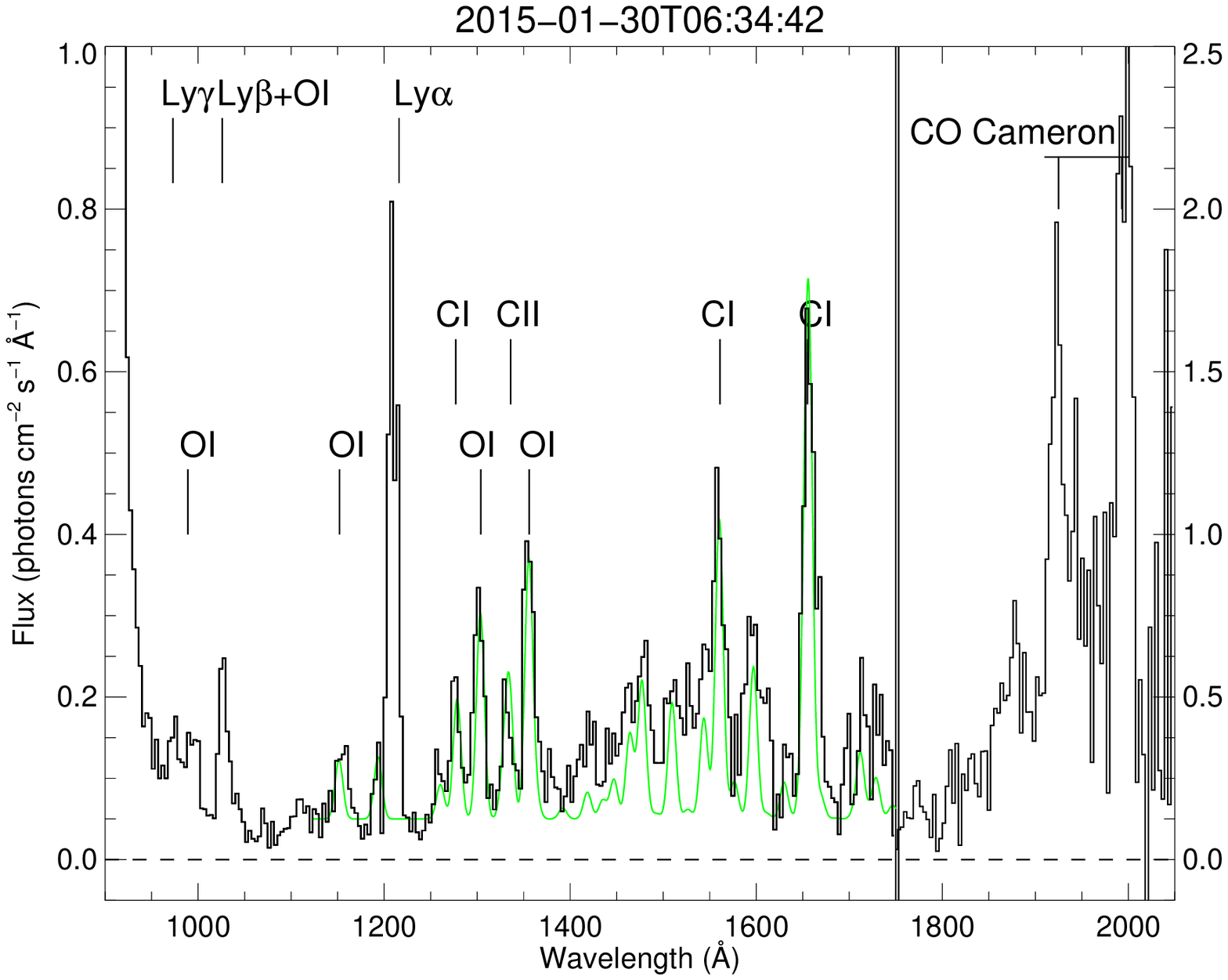}
\includegraphics*[width=0.44\textwidth,angle=0.]{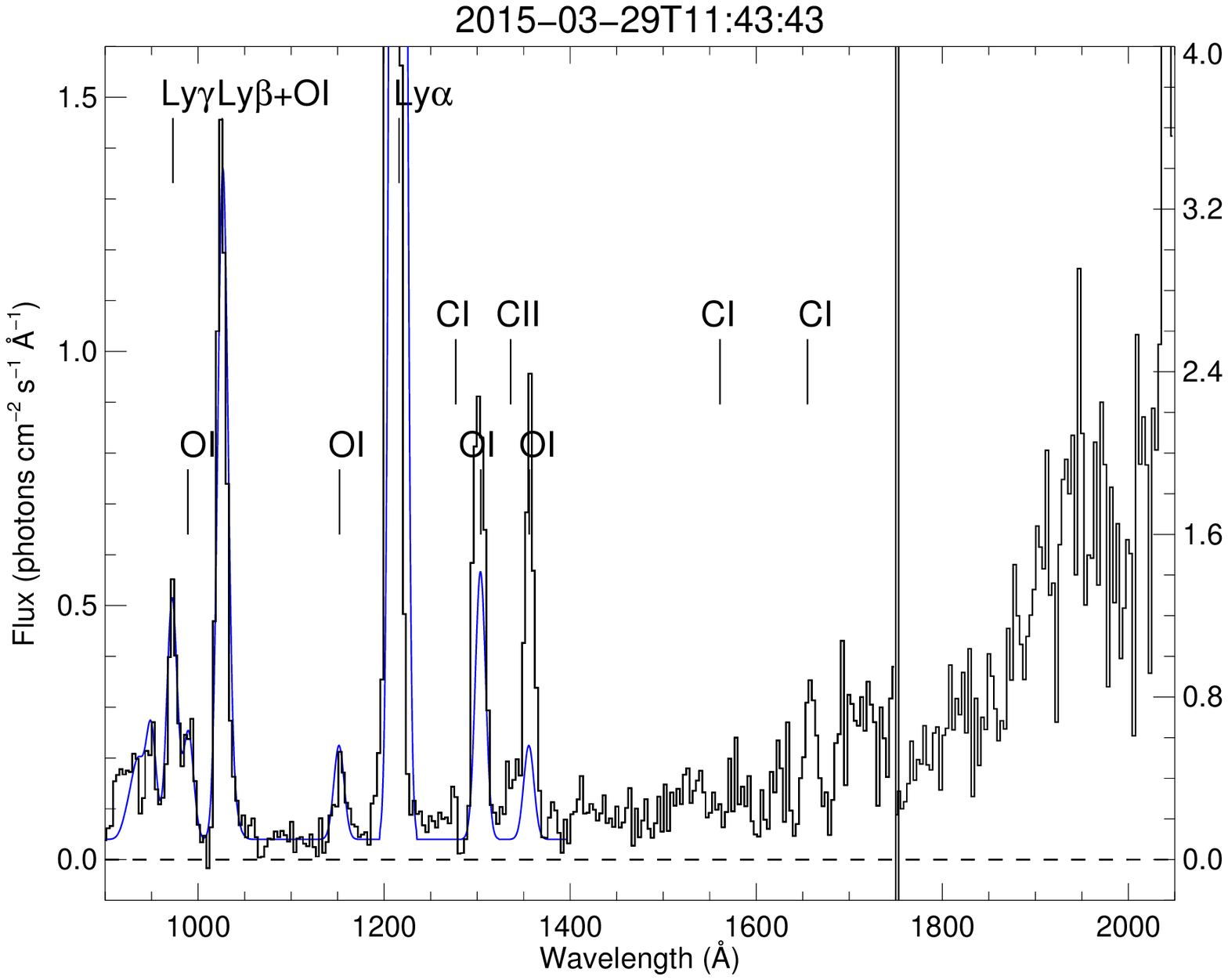}
\caption{Coma spectra corresponding to the spectral images in Figure~\ref{nav1}.  All of the spectra are summations over 4 rows (0.05\degr\ $\times$ 1.2\degr) in the narrow center of the slit.  The blue line is a synthetic spectrum of  electron impact on \Htwoo.  The green line is the same for \cotwo.  Both are adjusted to compensate for uncertainties in the energy distribution of both the cross section and electron flux (see text).  The positions of the (2,0) and (1,0) CO Cameron bands are indicated on the middle panel.  Note the scale change at 1750~\AA\ by a factor of 2.5. 
\label{spec1}}
\end{center}
\end{figure}

The observations presented here are listed, together with the observing parameters, in Table~\ref{obs}.  The first three are pre-perihelion; the last two post-perihelion.  The 2014 November 29 observation was discussed by \citet{Feldman:2015}, but the spectra presented here use an up-to-date instrument calibration.  The pre-perihelion spectral images are shown in Fig.~\ref{nav1}, each with a near-simultaneous context image from one of {\it Rosetta's} navigation cameras (NAVCAM) on which the location of the Alice slit is superimposed.
The corresponding spectra are shown in Fig.~\ref{spec1}.  Similarly, the post-perihelion spectral images and spectra are shown in Figures~\ref{nav2} and \ref{spec2}.

\begin{deluxetable*}{lccccccc}
\tabletypesize{\small}
\tablewidth{0pt}
\tablecolumns{8}
\tablecaption{Log of Alice Observations. \label{obs}}
\tablehead{
\colhead{Date} & \colhead{Start} & \colhead{$r_h$\tablenotemark{a}} & \colhead{$d$\tablenotemark{b}} & \multicolumn{2}{c}{Sub-spacecraft} &  \colhead{Phase} & \colhead{Integration} \\
\colhead{} & \colhead{Time (UT)} & \colhead{(AU)} & \colhead{(km)} & \colhead{Longitude\tablenotemark{c} (\arcdeg)} &  \colhead{Latitude\tablenotemark{c} (\arcdeg)} &  \colhead{Angle\tablenotemark{c} (\arcdeg)} & \colhead{Time (s)} }
\startdata
\multicolumn{8}{l}{Observations Against the Shadowed Nucleus}  \\
2014 Nov 29 & 18:00 & 2.87 & 30.3 & 180 & 51.5 &  93.1 & 2419 \\
2015 Jan 30 & 06:34 & 2.43 & 27.8 & --12 & --64.2 &  92.0 & 2902 \\
2015 Mar 29 & 11:43 & 1.99 & 92.3 & --93 & 7.5 &  83.9 & 1209 \\ 
\hline
2016 Apr 26 & 06:06 & 2.88 & 21.2 & --107 & --33.1 &  109.1 & 4452 \\
2016 May 14 & 14:39 & 3.00 & 9.8 & --111 & --53.1 &  89.2 & 4613 \\
\hline
\multicolumn{8}{l}{Limb Observations}  \\
2015 Aug 19 & 15:29 & 1.25 & 329.8 & 152.8 & 24.3 &  89.7 & 11410 \\
2016 Sep 25 & 17:24 & 3.81 & 16.8 & --66.5 & --73.6 &  114.4 & 1821 \\
\enddata
\vspace*{-2pt}
\tablenotetext{a}{Heliocentric distance.}
\tablenotetext{b}{Distance to center of the comet.}
\tablenotetext{c}{At start time of histogram integration.}
\end{deluxetable*}

\subsubsection{2014 November 29}

As noted above, since the distant coma and IPM background are blocked by the nucleus, the observed ratio of Lyman-$\beta$ to Lyman-$\alpha$ along the line-of-sight confirms the interpretation as due to electron impact dissociation of \Htwoo\ based on laboratory cross section measurements at 200~eV of \citet{Makarov:2004}.  A synthetic spectrum based on these cross sections is shown in the top panel of Fig.~\ref{spec1}.  Only the \ion{O}{1} $\lambda$1152 cross section, which shows a different energy dependence from \ion{O}{1} $\lambda$1304 by \citeauthor{Makarov:2004},
is reduced by a factor of two to match the Alice spectrum.  We note that the cross section for Lyman-$\beta$ given by \citeauthor{Makarov:2004} includes the nearly coincident \ion{O}{1} $\lambda$1025.72 line that is not resolved from Lyman-$\beta$ in their experiment.  These lines are also blended in the Alice spectra and in the following discussion Lyman-$\beta$ will denote the combined emission feature.

The measured brightnesses of the strongest emissions (excluding Lyman-$\alpha$) are given in Table~\ref{data}.  The absence of \ion{C}{1} emissions indicates a very low relative abundance of \cotwo\ to \Htwoo, consistent with the ROSINA/RTOF measurements \citep{Mall:2016} that show that \Htwoo\ is the dominant coma species at northern latitudes during this time frame (pre-equinox).  As we had noted previously \citep{Feldman:2015}, photodissociation of \Htwoo\ was found to be too slow a process \citep{Wu:1993} to produce the observed emissions, so we take this spectrum as a ``signature" of dissociative electron excitation of \Htwoo\ in the coma.  Due to uncertainties in both electron energy distribution and gas density along the line-of-sight we do not attempt to fully model these emissions but solely to use the observed line ratios to identify the source of the emission.

\subsubsection{2015 January 30 \label{jan}}

\citet{Mall:2016}, and also \citet{Hassig:2015}, from earlier data, show that at extreme southern latitudes, pre-equinox, the ratio of \cotwo\ to \Htwoo\ can often exceed 2.  Alice observations toward the shadowed nucleus at southern latitudes on 2015 January 30 show a ``picket fence" pattern of carbon and oxygen multiplets \ion{C}{1} $\lambda$1277, \ion{O}{1} $\lambda$1304, \ion{C}{2} $\lambda$1335, and \ion{O}{1} $\lambda$1356 (and weak \ion{C}{1} $\lambda$1260), together with bands of the CO Fourth Positive system.  These were the earliest spectra showing coma emissions from both \ion{C}{1} and CO that are stronger than the \ion{H}{1} and \ion{O}{1} emissions from the dissociation of \Htwoo, indicative of  a higher column abundance of either \cotwo\ or CO.  At the long-wavelength end of the Alice spectral range, where the instrument sensitivity is decreasing rapidly, CO Cameron bands, particularly the (3,0), (2,0), and (1,0) bands at 1868, 1928, and 1993~\AA, respectively, are clearly seen in the middle spectral image of Fig.~\ref{nav1}. These bands, similar to those seen in the spectrum of Mars that are characterized by a rotational temperature $\geq 1000$~K \citep{Conway:1981,Jain:2015}, are primarily produced by dissociative excitation of \cotwo\ \citep{Weaver:1994}, and will be discussed in a separate paper.

Laboratory spectra of \citet{Mumma:1971} show that in electron excitation of \cotwo\ the $v' = 0$ and $v' = 1$ levels of the CO \api\ state are comparably populated while for electron excitation of CO $v' = 0$ is a factor of two lower than $v' = 1$.  More importantly, \citet{Ajello:1971a} showed that electron impact on CO would produce emission of \ion{C}{2} $\lambda$1335 much stronger than any of the neighboring atomic C or O multiplets. Thus, the observed spectra suggest 
that the ``picket fence" results from electron-impact dissociative excitation of \cotwo\ and not direct excitation of CO.  This spectrum is prevalent when the comet is far from the Sun when {\it Rosetta} is close to the nucleus and when \cotwo\ is abundant relative to \Htwoo, i.e., at southern latitudes.
Then contributions of \Htwoo\ and \otwo\ to the \ion{O}{1} emissions are minimal and the ``picket fence" may be used as a ``signature" of electron impact dissociative excitation of \cotwo.  Such a synthetic spectrum, based on laboratory cross sections of \citet{Mumma:1971}, \citet{Ajello:1971a}, \citet{Wells:1972b}, and \citet{Kanik:1993}, all normalized to 100~eV, is shown in the middle panel of Fig.~\ref{spec1}.  Since the electron flux energy distribution is known to be variable, both in time and with distance to the nucleus \citep{Madanian:2016}, and not all cross sections have measured energy dependence, adjustments of up to 25\%s are made to match the synthetic spectrum to the observed Alice spectrum.

In our earlier report \citep{Feldman:2015}, we used the measured ratio of \ion{C}{1} $\lambda$1657 to Lyman-$\beta$, using the ratio of excitation cross sections at 100~eV of 0.62, to determine the relative abundance of \cotwo\ to \Htwoo\ along the line-of-sight.  We add the caveat that the observed \ion{C}{1} $\lambda$1657 emission also includes the unresolved (0,2) Fourth Positive band at 1653~\AA, that can be subtracted using the nearby unblended (0,1) band at 1597~\AA\ and the tabulated branching ratios of \citet{Morton:1994}.  For this spectrum, using the values in Table~\ref{data}, we find a \ion{C}{1} $\lambda$1657/Lyman-$\beta$ ratio of $4.1 \pm 1.6$, which translates to an abundance ratio [\cotwo]/[\Htwoo] $= 2.6 \pm 1.0$.  This value is consistent with the maximum {\it in situ} values found by \citet{Mall:2016} at extreme southern latitudes, although we note that they do not report such a high value for this particular day.

\subsubsection{2015 March 29}

Following a close-fly-by of the nucleus on 2015 March 28, the {\it Rosetta} star trackers became confused by scattered light from dust and the spacecraft went into a safe mode at UT 12:15 on 2015 March 29.  After the recovery, the orbit distance was increased to $>100$~km as a safety measure which made it not possible with the spatial resolution of the Alice slit to resolve shadowed regions of the nucleus until the orbit distance was again reduced several months after perihelion.  The final spectrum with the narrow center of the Alice slit centered on the shadowed region of the nucleus was obtained beginning 30 minutes before the safing and is shown on the lower panels of Figs.~\ref{nav1} and \ref{spec1}.

\begin{figure*}[ht]
\begin{center}
\includegraphics*[width=0.37\textwidth,angle=0.]{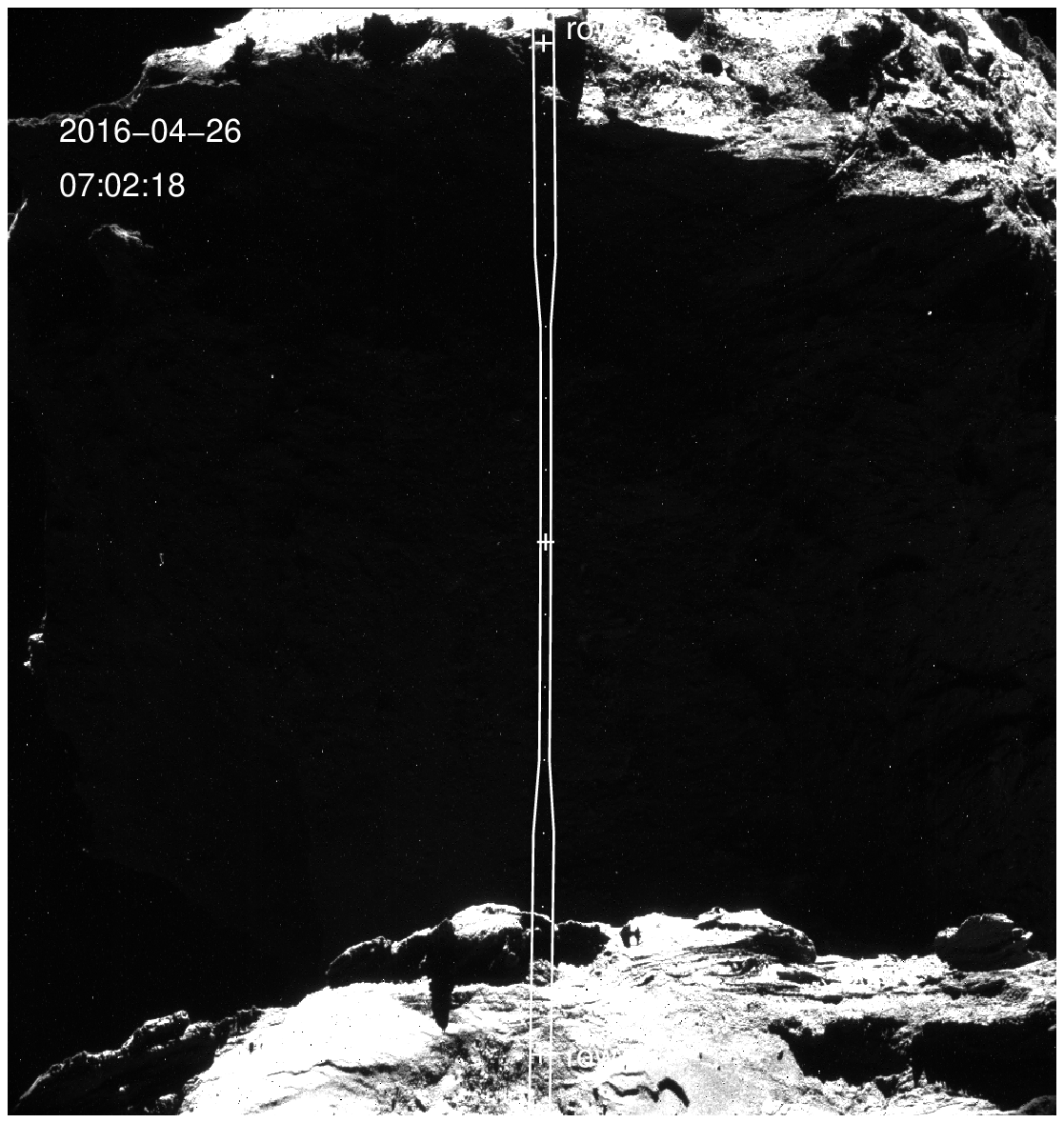}
\includegraphics*[width=0.62\textwidth,angle=0.]{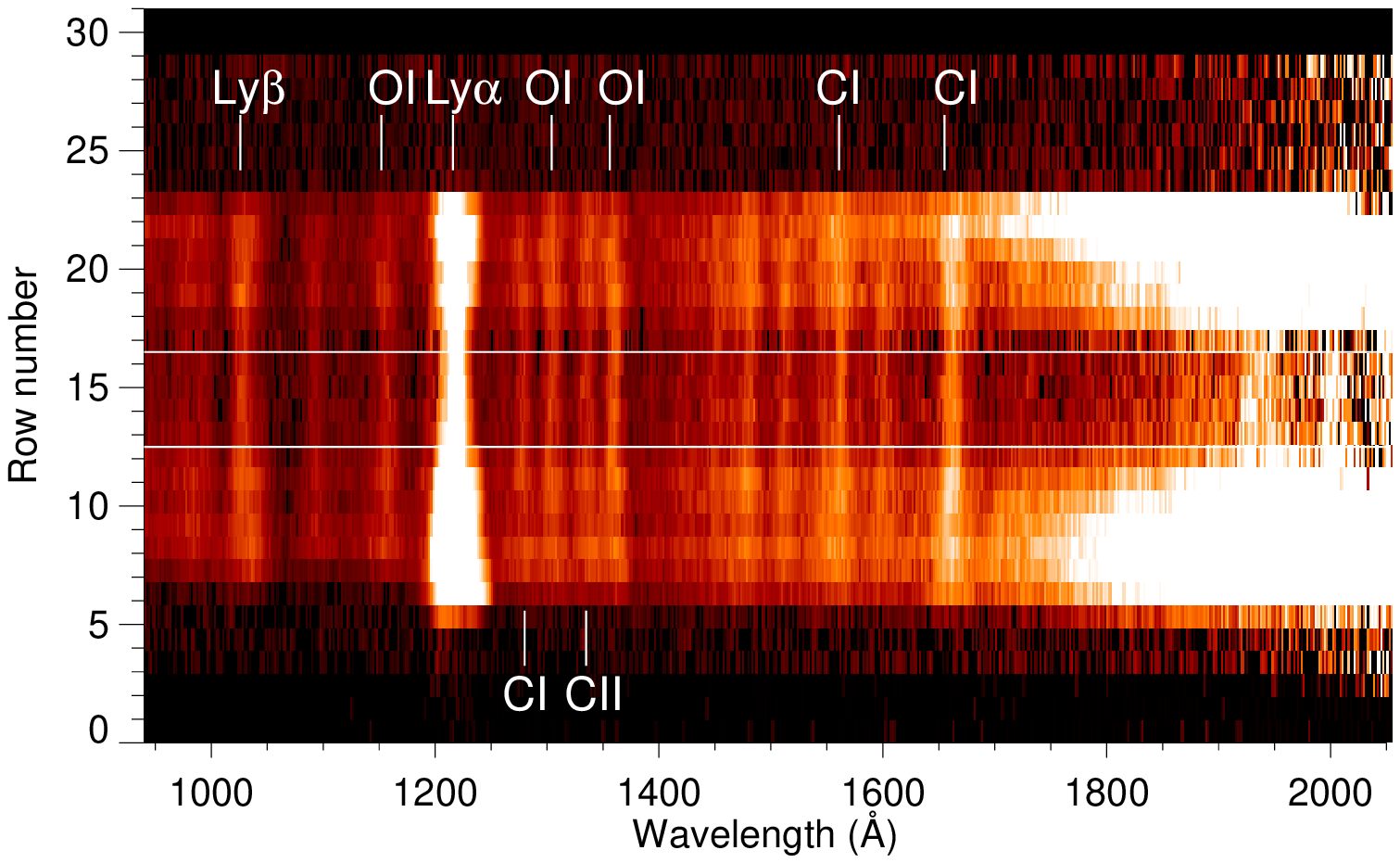}
\includegraphics*[width=0.37\textwidth,angle=0.]{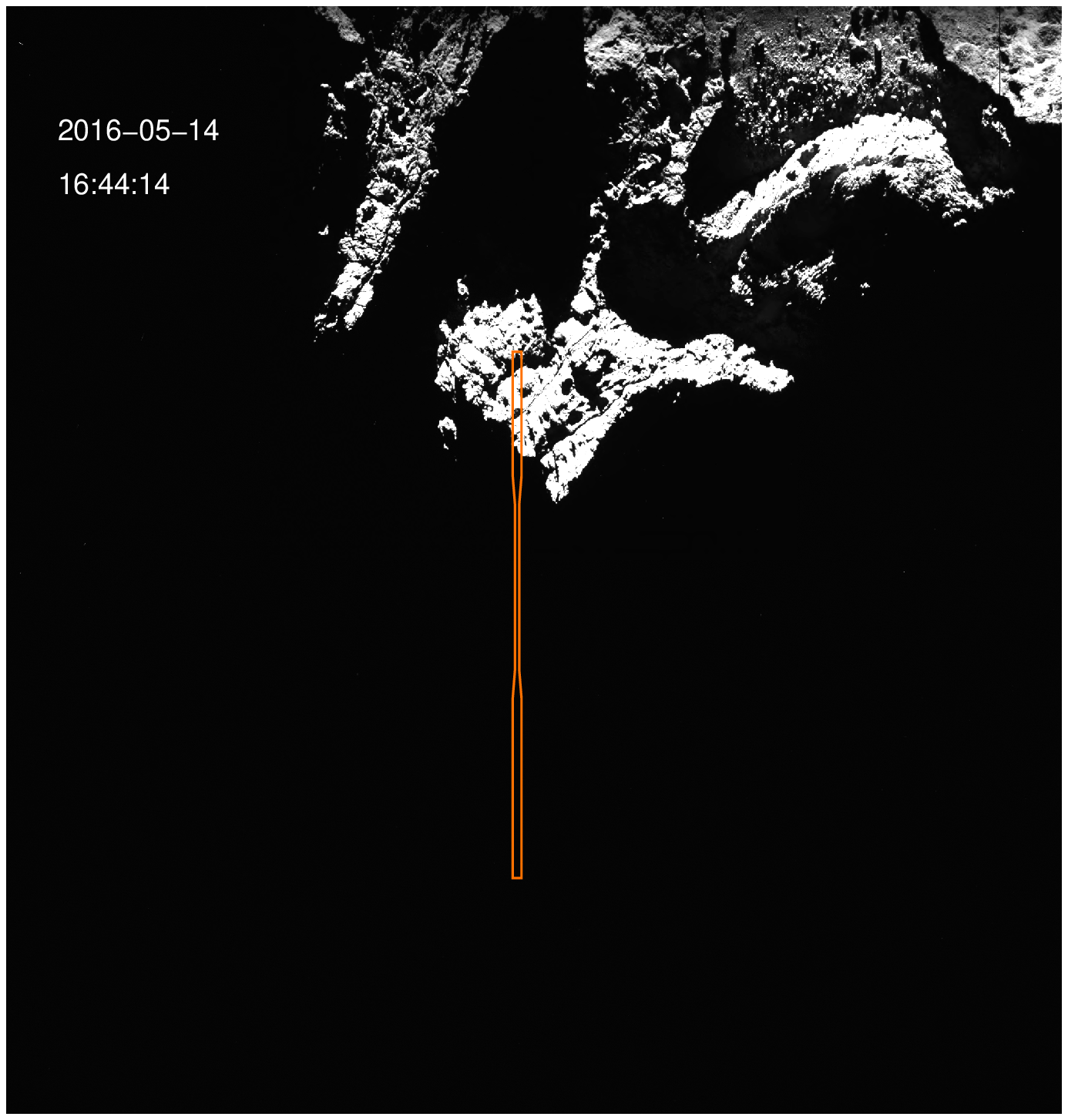}
\includegraphics*[width=0.62\textwidth,angle=0.]{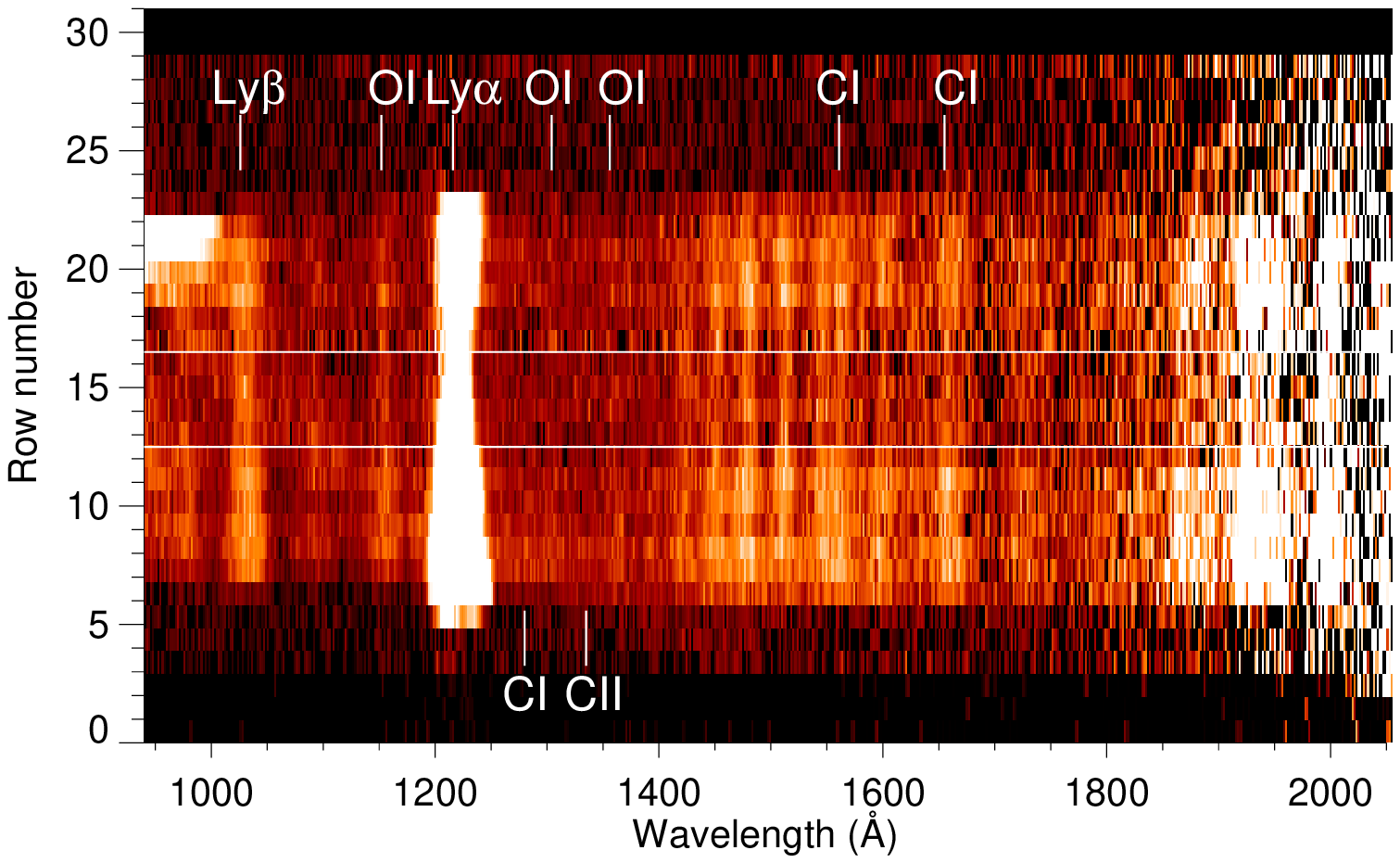}
\caption{Same as Figure~\ref{nav1} for the two post-perihelion dates given in Table~\ref{obs}, except that the context image for 2016 May 14 is from the OSIRIS wide angle camera.   \label{nav2} }
\end{center}
\end{figure*} 

\begin{deluxetable*}{lccccccc}
\tabletypesize{\small}
\tablewidth{0pt}
\tablecolumns{6}
\tablecaption{Observed Muliplet Brightnesses (rayleighs). \label{data}}
\tablehead{
\colhead{Date} & \colhead{Start Time (UT)} & \colhead{\ion{H}{1} Lyman-$\beta$} & \colhead{\ion{O}{1} $\lambda$1304} & \colhead{\ion{O}{1} $\lambda$1356} & \colhead{\ion{C}{1} $\lambda$1657}  }
\startdata
2014 Nov 29 & 18:00 & $22.3 \pm 2.6$ & $7.93 \pm 0.94$ & $2.73 \pm 0.60$ & $0.89 \pm 1.02$ \\
2015 Jan 30 & 06:34 & $1.35 \pm 0.31$ & $1.91 \pm 0.43$ & $2.85 \pm 0.58$ & $5.95 \pm 1.24$ \\
2015 Mar 29 & 11:43 & $12.3 \pm 2.1$ & $9.18 \pm 1.96$ & $7.24 \pm 1.63$ & $3.11 \pm 1.93$ \\
\hline
2016 Apr 26 & 06:06 & $4.58 \pm 0.88$ & $3.93 \pm 0.66$ & $5.86 \pm 1.09$ & $12.6 \pm 1.4$ \\
2016 May 14 & 14:39 & $3.07 \pm 0.31$ & $-0.47 \pm 0.46$ & $0.01 \pm 0.42$ & $1.90 \pm 0.64$ \\
\hline
\enddata
\end{deluxetable*}

The spectrum is again different from the two prior examples.  The \ion{O}{1} $\lambda$1304 and \ion{O}{1} $\lambda$1356 multiplets are of comparable brightness and higher than expected relative to Lyman-$\beta$ for dissociative electron excitation of \Htwoo.  This behavior was noted in several instances of gaseous outbursts and attributed to electron dissociative excitation of \otwo\ \citep{Feldman:2016}.  Subtracting an \Htwoo\ synthetic spectrum leaves the $\lambda$1356/$\lambda$1304 ratio $\approx 2$, as expected for electron impact on \otwo\ \citep{Kanik:2003}.  Assuming relatively similar energy dependence on the excitation cross sections \citep{Kanik:2003,Makarov:2004}, and no contribution from dissociative excitation of \cotwo, we can then use the observed \ion{O}{1} $\lambda$1356/Lyman-$\beta$ ratio to determine the abundance of \otwo\ relative to \Htwoo\ along the line-of-sight to the nucleus, which in this case is 0.07.  This is probably slightly underestimated as electron excitation of \otwo\ will also contribute to the Lyman-$\beta$ + \ion{O}{1} $\lambda$1025.72 blend \citep{Ajello:1985}.
The relative \otwo\ abundance is somewhat higher than the mean value derived from ROSINA measurements \citep{Bieler:2015}, but not inconsistent with absorption measurements made by Alice \citep{Keeney:2017}.  The viewing geometry confirms that the observed \otwo\ is emitted directly from the nucleus.  We note that this \otwo\ ``signature" first appeared in Alice spectra beginning in late February 2015 and in addition to being seen in several short outbursts, as noted above, continued to be seen regularly through the following February.

\subsubsection{Post-perihelion, post-equinox, 2016 April 26  \label{post}}

Post-perihelion and post-equinox (2016 March 23), as the orbit distance to the comet was reduced we could once again observe against the shadowed nucleus.  The ``picket fence" spectrum appeared frequently during the last six months of the mission, up to the last limb observation made on 2016 September 29.  At the same time, atomic emissions from the dissociation of \Htwoo\ became weaker suggesting that water outgassing was decreasing much more rapidly as the comet receded from the Sun than was \cotwo\ \citep{Hansen:2016}.  The outgassing of both \Htwoo\ and \cotwo\ thus appears asymmetric about perihelion.  An example is the 2016 April 26 observation listed in Table~\ref{obs}.  The spectral image and spectrum are shown in Figs.~\ref{nav2} and \ref{spec2}.  In addition to electron excited emission, CO Fourth Positive fluorescence is also present as indicated by the red synthetic spectrum based on the model of \citet{Lupu:2007}, with rotational temperature (75~K) and outflow velocity (0.7~\kms) typical of comets near 1~AU.  As we did above, we can use the observed \ion{C}{1} $\lambda$1657/Lyman-$\beta$ ratio to determine the relative \cotwo\ abundance along the line-of-sight.  In this case we must also take into account the contribution to \ion{C}{1} $\lambda$1657 from the (0,2) band excited by resonance fluorescence.  The result is  [\cotwo]/[\Htwoo] $= 1.1 \pm 0.1$.

\subsubsection{2016 May 14  \label{post2}}

A particularly interesting case is the 2016 May 14 observation, in which both the \Htwoo\ and \cotwo\ signatures appear weak or absent while CO Fourth Positive emission is strong.  This suggests a concentrated source of CO at the sub-spacecraft point on the nucleus, as well as a very low flux of suprathermal electrons in the inner coma. In both of these spectra, emission from the CO \csig\ -- \xsig\ (0,0) Hopfield-Birge band at 1087~\AA, also excited by solar fluorescence \citep{Feldman:2002} is seen.

\begin{figure}[h]
\begin{center}
\includegraphics*[width=0.46\textwidth,angle=0.]{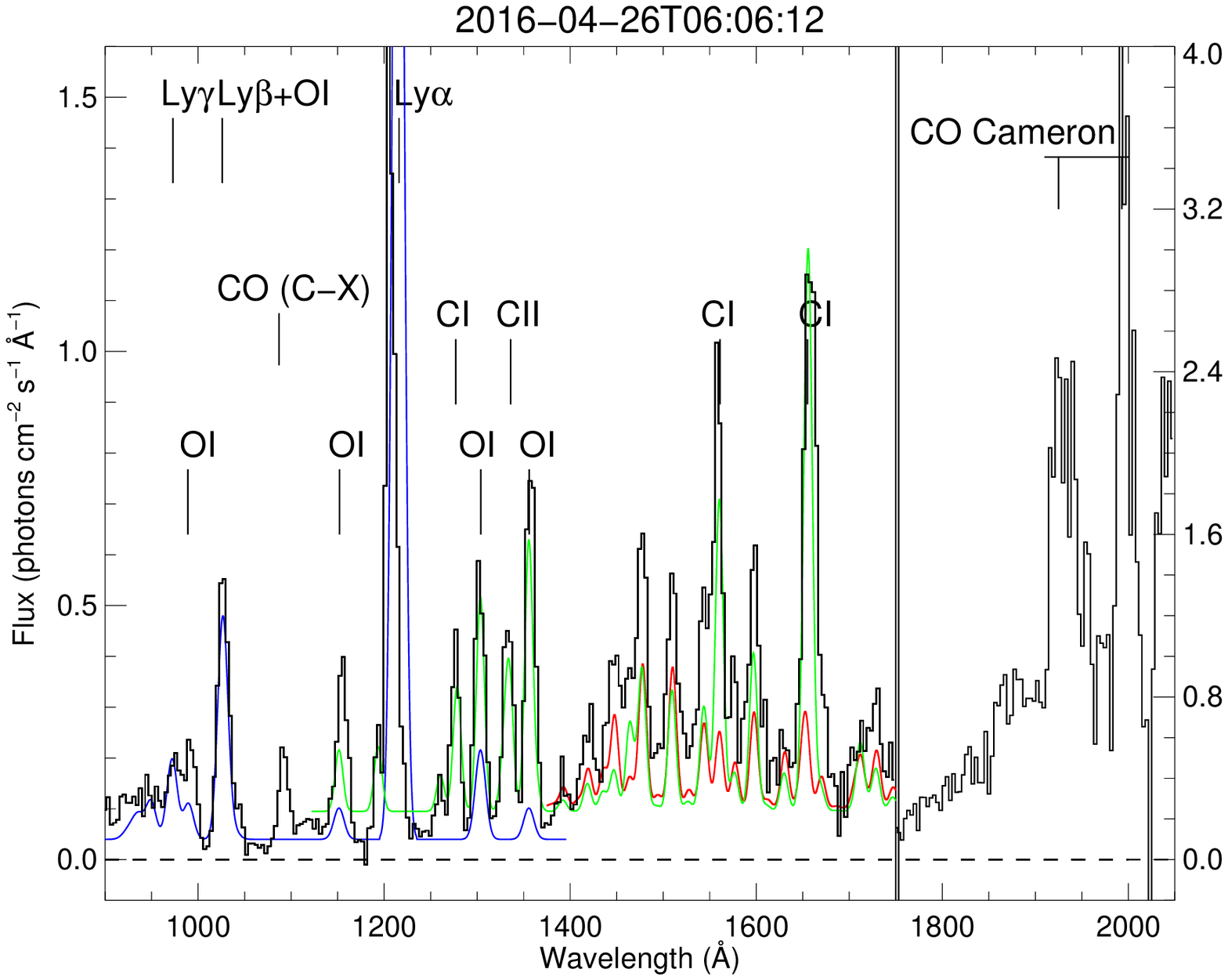}
\includegraphics*[width=0.46\textwidth,angle=0.]{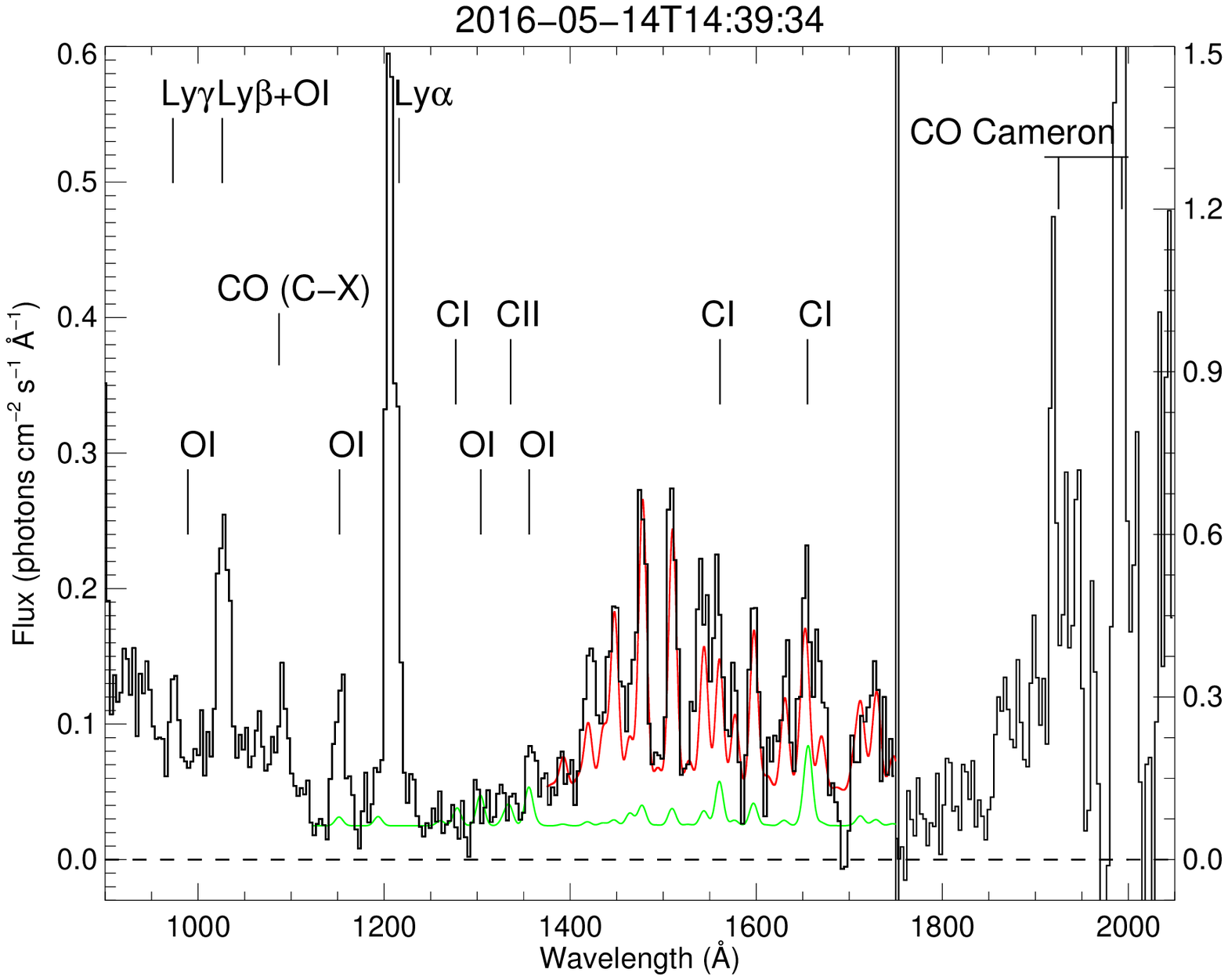}
\caption{Coma spectra corresponding to the spectral images in Figure~\ref{nav2}.  The blue line is a synthetic spectrum of  electron impact on \Htwoo.  The green line is the same for \cotwo.  The red line is a synthetic CO Fourth Positive fluorescence spectrum following \citet{Lupu:2007} for a CO column density of $1.6 \times 10^{14}$~cm$^{-2}$ (April 26) and $1.0 \times 10^{14}$~cm$^{-2}$ (May 14).  Note the distorted Lyman-$\alpha$ line shape due to detector degradation effects described in Section~\ref{inst}.  \label{spec2}}
\end{center}
\end{figure} 

\cotwo\ was definitely present as evidenced by $\sim$20~rayleighs of (1,0) Cameron band emission at 1993~\AA, produced primarily by photodissociation. Taking the photodissociation rate at solar minimum for the (1,0) band from \citet{Feldman:1997} of $8.3 \times 10^{-8}$ photons~s$^{-1}$~molecule$^{-1}$ at 1~AU, we find a \cotwo\ column density of $\sim 2 \times 10^{15}$ cm$^{-2}$.  The fluorescence model of \citet{Lupu:2007} gives a CO column density of $1.0 \times 10^{14}$ cm$^{-2}$, and thus a CO/\cotwo\ ratio of $\sim$0.05.  The \ion{C}{1} $\lambda$1657 multiplet is also produced in photodissociation of \cotwo\ and \citet{Wu:1988} give an excitation rate of $\sim 1.0 \times 10^{-9}$ photons~s$^{-1}$~molecule$^{-1}$ at 1~AU.  This would give only $\sim$0.25~rayleighs of \ion{C}{1} $\lambda$1657, about one-tenth the amount observed in the lower panel of Figure~\ref{spec2}.  \citeauthor{Wu:1988} also note that the rate for photodissociation of CO to produce \ion{C}{1} $\lambda$1657 is an order of magnitude higher than for \cotwo, but the lower relative abundance of CO does not make this a significant additional source.  Thus the remainder of this emission must come from electron impact on \cotwo, as shown by the green synthetic spectrum in the figure.

\subsection{Limb observations around perihelion \label{limb}}

\begin{figure*}[ht]
\begin{center}
\includegraphics*[width=0.37\textwidth,angle=0.]{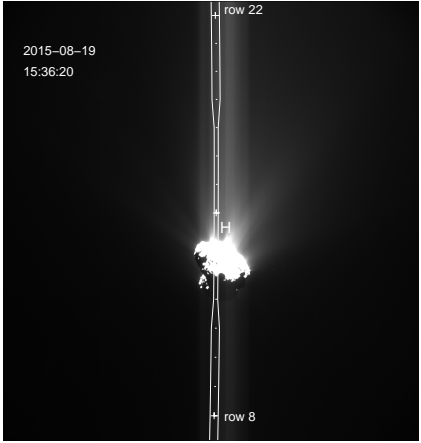}
\includegraphics*[width=0.62\textwidth,angle=0.]{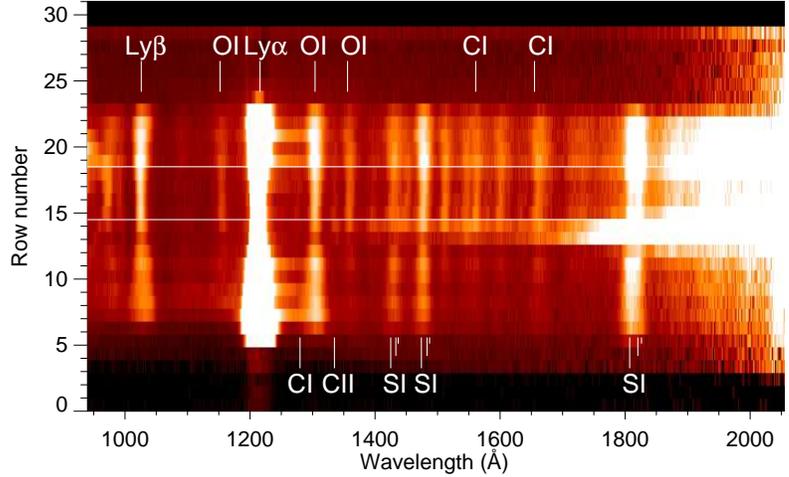}
\caption{Same as Figure~\ref{nav1} for 2015 August 19.  The heliocentric distance was 1.246~AU, the distance of {\it Rosetta} to the center of the comet was 330~km, and the solar phase angle was 89.7\degr.  The approximate position of the VIRTIS-H slit is marked by ``H'' in the NAVCAM image.  The Alice spectral image is a co-addition of 20 histograms for a total integration time of 11,410~s.  \label{aug_image} }
\end{center}
\end{figure*}

\begin{figure}[ht]
\begin{center}
\includegraphics*[width=0.49\textwidth,angle=0.]{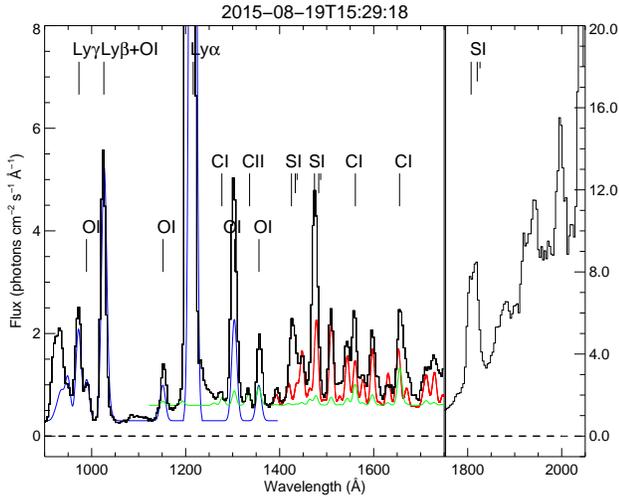}
\caption{Coma spectrum corresponding to the spectral image in Figure~\ref{aug_image}.  The blue line is a synthetic spectrum of  electron impact on \Htwoo.  The green line is the same for \cotwo.  The red line is a synthetic CO Fourth Positive fluorescence spectrum following \citet{Lupu:2007} for a CO column density of $1.8 \times 10^{14}$~cm$^{-2}$.  \label{aug_spec}}
\end{center}
\end{figure}

We can compare the spectra described above, which were made against the shadowed nucleus, to limb observations made in the several months around perihelion (2015 August 13.09) when {\it Rosetta} was $\geq$150 km from the comet.  NAVCAM images showed strong activity above the sunward limb and the Alice spectra showed a strong sunward-anti-sunward asymmetry in both the gas emissions and dust reflected sunlight.  Two examples are listed in Table~\ref{obs}.  A spectral image from 2015 August 19 is shown in Fig.~\ref{aug_image}.
The corresponding Alice spectrum, shown in Fig.~\ref{aug_spec} is unlike any of the examples discussed above.  CO Fourth Positive and Cameron bands are present but atomic emissions, other than those produced by resonance scattering, particularly \ion{O}{1} $\lambda$1304 and three \ion{S}{1} multiplets identified in the figure, are significantly weaker.  The atomic sulfur emissions are resonantly scattered radiation from the dissociation products of the many sulfur-containing molecules identified by ROSINA \citep{Calmonte:2016}, and will be discussed in a separate paper.  Like the \ion{S}{1} multiplets, both \ion{O}{1} $\lambda$1304 and Lyman-$\beta$ show emission from resonance scattering on the anti-sunward side of the nucleus from the extended coma, whose presence complicates the extraction of the electron excited emissions on the sunward side.

\begin{figure}[ht]
\begin{center}
\includegraphics*[width=0.5\textwidth,angle=0.]{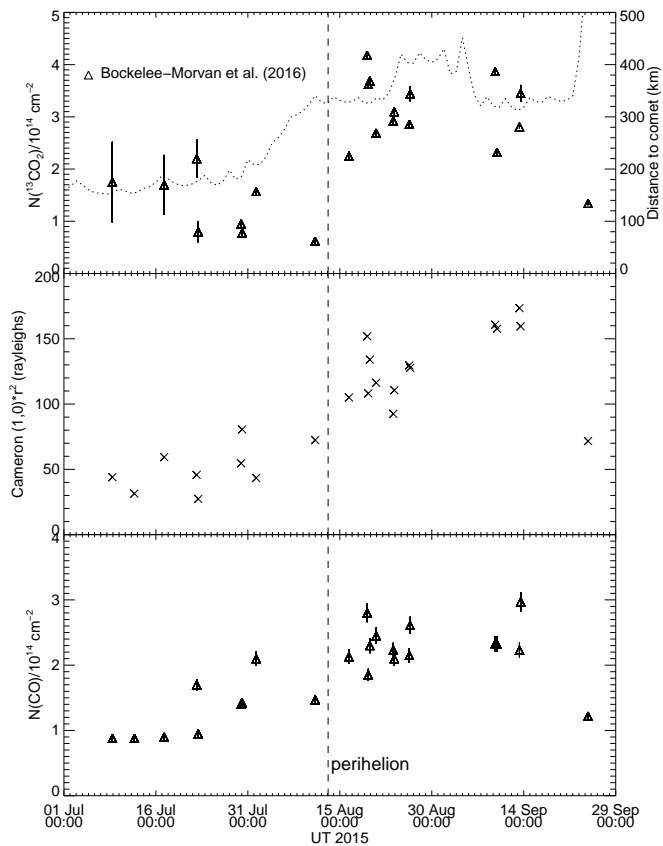}
\caption{Top: Derived $^{13}$\cotwo\ column densities from VIRTIS-H spectra from long limb stares \citep{Bockelee:2016}.  The dotted line shows the distance of {\it Rosetta} to the comet.  Middle: Brightness of the (1,0) Cameron band scaled to the square of the heliocentric distance.  Bottom: Derived CO column density assuming resonance fluorescence and the model of \citet{Lupu:2007}.  \label{dbm}}
\end{center}
\end{figure}

The CO Cameron bands lie at the long wavelength end of the Alice spectral range and are superimposed on a rising solar continuum of reflected light from the dust seen in the NAVCAM images.  They are consistent with laboratory measurements of photodissociatve excitation of \cotwo\ in which the bands appear narrower ($T_{rot} \sim 400$~K) than those produced by electron dissociative excitation ($T_{rot} \sim 1600$~K) \citep{Conway:1981}.  There are calibration issues in this part of the spectrum as seen the the ratio of the (2,0) to (1,0) bands which does not track the predicted value but may be due to the presence of the unresolved \ion{C}{1} $\lambda$1931 line.
Nevertheless, we can compare the Alice data to VIRTIS-H measurements of \cotwo\ column density made during a number of long (2--4 hours) off-limb stares made between 2015 July through September in which Alice observed simultaneously \citep{Bockelee:2016}.  The extracted brightness of the Cameron (1,0) band, despite the many uncertainties and scaled by $r_h^2$, tracks very closely with the VIRTIS-H measurement of \cotwo\ column density with time during this period as shown in Fig.~\ref{dbm}.  A quantitative comparison is given below in Section~\ref{vh}.
With a few exceptions, the Fourth Positive bands fit very well to the resonance fluorescence model of \citet{Lupu:2007}, and the derived column densities, shown in the lower panel of Fig.~\ref{dbm}, are also found to track the \cotwo\ column densities.

\subsection{Observations near the end of mission}

\begin{figure}[ht]
\begin{center}
\includegraphics*[width=0.5\textwidth,angle=0.]{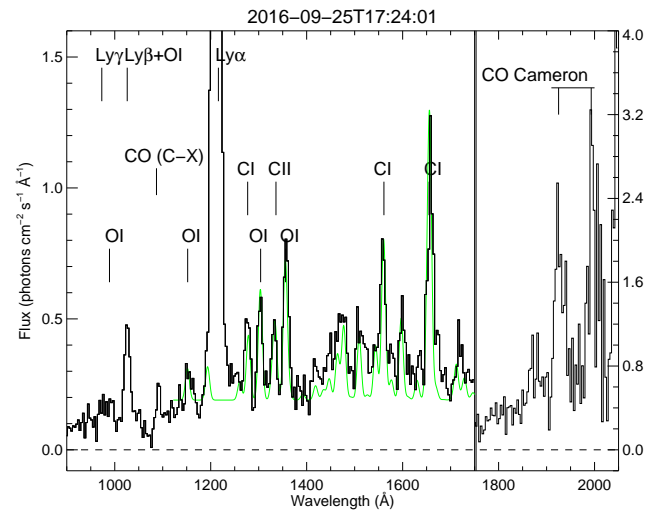}
\caption{Alice limb spectrum from 2016 September 25, five days before the end of the mission.  The observing parameters are listed in Table~\ref{obs}.  The spectrum is due almost entirely to dissociative electron excitation of \cotwo.  \label{sep_spec}}
\end{center}
\end{figure}

As the {\it Rosetta} spacecraft prepared to touch down on the nucleus on 2016 September 30, there was little opportunity for spectroscopic observations of either the nucleus in shadow or the near-nucleus coma.  The final off-limb spectrum was obtained by Alice at UT 02:16 on 2016 September 29.  It is extremely noisy but recognizable and quite similar to the spectrum obtained four days earlier which we show in Fig.~\ref{sep_spec}.  For this spectrum the limb is at the bottom of the Alice slit and the mean distance of the center of the slit from the limb is $\sim$0.5~km.  The sub-spacecraft latitude was --74\degr\ and the spectrum is almost identical to the electron excited \cotwo\ spectrum observed on 2015 January 30 (middle panel, Figure~\ref{spec1}), also at far southern latitude.  Again we can use the observed \ion{C}{1} $\lambda$1657/Lyman-$\beta$ ratio to determine the relative \cotwo\ abundance along the line-of-sight and this gives [\cotwo]/[\Htwoo] $= 1.8 \pm 0.6$.  However, this should be regarded as a lower limit as the Lyman-$\beta$ brightness may include up to half from resonance scattering in the coma, based on off-limb observations on 2016 September 27.

\section{DISCUSSION}

\subsection{Variability of electron excited emissions}

The Alice spectrograph obtained a large number of diverse far-ultraviolet spectra of the coma of comet 67P/Churyumov-Gerasimenko over the entire escort phase of the mission.  As we have demonstrated above, much of the observed emission, particularly for $r_h > 2$~AU is produced by electron impact dissociative excitation of primary coma molecules, \Htwoo, \cotwo, and \otwo, close to the nucleus.  While the electron flux at the spacecraft can be determined from {\it in situ} measurements by the various instruments of the Rosetta Plasma Consortium, the variation of the excitation along the Alice line-of-sight is not known {\it a priori}, so we cannot directly extract molecular column densities from the spectra, only relative abundances.  Variability of the observed emissions is thus due to both the coma gas density at a given position relative to the nucleus, but also the plasma environment at that time and position.  An example of the correlation between the two can be found from a comparison of the Alice spectra from 2014 October 18-19 \citep{Feldman:2015} with electron flux models for the same dates based on RPC-LAP data from \citet{Galand:2016}.  The Alice data were taken during off-nadir limb stares that coincided with the interval ``T2'' discussed by \citeauthor{Galand:2016} in their Figures 15 and 16.  The variation in observed Lyman-$\beta$ brightness closely follows that of the electron model, both showing a sharp maximum at UT 00:00 on 2014 October 19.

\subsection{Comparison with coma models \label{models}}
\setcounter{footnote}{0}

A number of models have been developed to describe the heterogeneous distribution of volatiles in the nucleus using the {\it in situ} mass spectrometer measurements from the ROSINA instruments \citep{Fougere:2016,Hansen:2016,Kramer:2017}, together with shape and illumination models. The DSMC models of \citet{Fougere:2016} are accessible through the ICES web site at the University of Michigan\footnote{\tt http://ices.engin.umich.edu}, which additionally provide integrated column densities along the Alice field-of-view using reconstructed trajectory and spacecraft NAIF/SPICE kernels.  As noted above, we cannot extract absolute column densities from observations of the electron excited emissions, but we can compare the relative abundances derived above with the predictions of the models, and we do so for the three pre-perihelion observations listed in Table~\ref{obs}, averaging the model over the narrow center of the Alice slit.  For 2014 November 29 the model predicts [\cotwo]/[\Htwoo] $= 0.035$, consistent with the very weak \ion{C}{1} $\lambda$1657 seen in Figure~\ref{spec1}, similar to what was found by \citet{Feldman:2015} for 2014 October 23.  For 2015 January 30 the model predicts a much higher value of [\cotwo]/[\Htwoo] $= 0.6$, but lower by a factor of 4--5 than found above.  This discrepancy was also noted by \citet{Fougere:2016} for the near-perihelion VIRTIS-H observations reported by \citet{Bockelee:2016}  The model ratio is consistent with that found by \citet{Mall:2016} for this time period but it should be noted that the model is based on the same dataset presented by \citeauthor{Mall:2016}  Finally, for 2015 March 29, the model ratio of [\otwo]/[\Htwoo] $= 0.07$ is in perfect agreement with the value deduced from the Alice spectrum above.

\subsection{Comparison with VIRTIS-H \label{vh}}

The position of the small VIRTIS-H slit relative to the long Alice slit is indicated by ``H''  in the NAVCAM image in Figure~\ref{aug_image}.  To compare the emissions seen by the two instruments requires the use of a spatial model of the gas emissions in the coma, and for this purpose we again turn to the DSMC models of \citet{Fougere:2016}.  \citeauthor{Fougere:2016} compare their model line-of-sight integrations with the VIRTIS-H measurements discussed in Section~\ref{limb}.  They find good agreement with the observed \cotwo\ column densities (assuming \cotwo/$^{13}$\cotwo\ = 89), but overestimate the \Htwoo\ column densities by a factor of $\sim$4.  From a model for UT 16:00 on 2015 August 19 we find that the geometric factor between the narrow center of the Alice slit and the VIRTIS-H aperture is 0.28 for the \cotwo\ column density.  Using the results in Table~3 of \citet{Bockelee:2016}, and the photoexcitation rate of the (1,0) Cameron band given in Section~\ref{post}, we would then expect Alice to observe 480 rayleighs of (1,0) band emission.  From the spectrum of Figure~\ref{aug_spec}, subtracting the dust scattered solar radiation, we find only 70 rayleighs.  The time period shortly after perihelion was the time of maximum activity and the highest \Htwoo\ column tabulated by \citeauthor{Bockelee:2016}  The photodissociation of \cotwo\ to produce CO Cameron bands peaks between 900 and 1000~\AA\ \citep{Lawrence:1972} and the cross section for \Htwoo\ absorption reaches a maximum of $\sim 2 \times 10^{-17}$~cm$^{2}$ in this same wavelength band \citep{Watanabe:1964}.  With a column density of $1.2 \times 10^{17}$~cm$^{-2}$ from \citeauthor{Bockelee:2016}, assuming a similar column along the line-of-sight to the Sun, we find an optical depth of $\sim 2$, which would be sufficient to reduce the (1,0) band excitation to the observed level.  The problem is in fact more complicated in that \cotwo\ and \Htwoo\ are mixed along the line-of-sight to the Sun, requiring a more detailed calculation.  However, the spatial distribution of the (1,0) band emission along the slit, relative to that of CO fluorescence, is relatively uniform, consistent with this behavior although it is also possible that there is significant collisional quenching of CO (\atpi) molecules in the innermost coma.  The derived abundance of CO relative to \Htwoo\ is $\sim 0.5$\%, consistent with the limit of \citeauthor{Bockelee:2016} of $\leq$1\%.  While the DSMC models overpredict both the CO and \Htwoo\ column densities, the relative abundance is in accord with the observations.

\section{SUMMARY}

The identification of distinct spectral signatures arising from multiple dissociative excitation processes of \Htwoo, \cotwo, \otwo, and CO, allows Alice to achieve one of its primary science objectives, the study of the evolution of the gaseous coma of comet 67P/Churyumov-Gerasimenko over the entire escort phase of the {\it Rosetta} mission.  Although our initial report started with data from 2014 September 21, emission due to electron impact on \Htwoo\ was first detected by Alice on 2014 August 18 when 67P was 3.53~AU from the Sun and {\it Rosetta} was $\sim$85~km from the comet.  The spatial distribution along the slit showed that the emission was concentrated within a few~km of the limb of the nucleus.  This pattern was repeated throughout the first eight months of the escort phase.  \cotwo, also excited by electron impact, was not detected until late 2015 January (Section~\ref{jan}), even at extreme southern latitudes where ROSINA data show [\cotwo]/[\Htwoo]$\sim$2 during the same time period \citep{Mall:2016}.  This apparent discrepancy is understood by the fact that both \Htwoo\ and \cotwo\ were below the detection limit of Alice at these times, \citeauthor{Mall:2016} showing that the \Htwoo\ density measured by ROSINA was much lower than at other points in the orbit of {\it Rosetta}.  The emissions observed by Alice also depend on the energetic electron flux near the nucleus which, in turn, depends on several variables including the neutral gas density \citep{Galand:2016}.

Similar viewing geometries were obtained during the final eight months of the escort phase as the comet receded from the Sun.  During this period the observed spectrum was dominated by electron impact excitation of \cotwo, which as noted above, was detected in the final off-limb spectrum obtained at a heliocentric distance of 3.84~AU on 2016 September 29, one day before the end of the mission.  The dominance of \cotwo\ over \Htwoo\ during this period is in accord with the ROSINA measurements of \citet{Gasc:2017}.  Qualitatively, the Alice observations of gas emission close to the nucleus give the same picture of the evolution of the \Htwoo\ and \cotwo\ coma as the {\it in situ} mass spectrometer measurements.

\acknowledgments

{\it Rosetta} is an ESA mission with contributions from its member states and NASA.  We thank the members of the {\it Rosetta} Science Ground System and Mission Operations Center teams, in particular Richard Moissl and Michael K\"uppers, for their expert and dedicated help in planning and executing the Alice observations.  We thank Andr\'e Bieler and Nicolas Fougere for their help and advice using the ICES models.  The Alice team acknowledges continuing support from NASA's Jet Propulsion Laboratory through contract 1336850 to the Southwest Research Institute.  The work at Johns Hopkins University was supported by a subcontract from Southwest Research Institute.

\vspace{5mm}
\facility{Rosetta}


\newpage

\listofchanges

\end{document}